\documentstyle[aps]{revtex}
\begin{document}
\def\Paris{Par\'\i{}s}
\def\Perez{P\'erez}
\def\Astrofisica{Astrof\'\i{}sica}
\def\Fisica{F\'\i{}sica}
\title{
Effective Action for Stochastic Partial Differential Equations
}
\author{
David Hochberg$^{+,*}$, Carmen Molina--\Paris$^{++}$,
Juan \Perez--Mercader$^{+++,*}$, and Matt Visser$^{++++}$
}
\bigskip
\address{ 
$^{+,+++}$Laboratorio de \Astrofisica\ Espacial y  \Fisica\
Fundamental, Apartado 50727, 28080 Madrid, Spain\\
$^{++}$Theoretical Division, Los Alamos National Laboratory,
Los Alamos, New Mexico 87545, USA\\
$^{++++}$Physics Department, Washington University,
Saint Louis, Missouri 63130-4899, USA\\
$^{*}$Centro de Astrobiolog\'\i{}a 
(Associate Member of NASA Astrobiology Institute), 
INTA, Ctra. Ajalvir, Km. 4, 28850 Torrej\'on, Madrid, Spain}
\bigskip
\date{5 April 1999; Cosmetic revisions 26 July 1999; \LaTeX-ed \today}
\maketitle

\bigskip

{\small

{\bf Abstract:} Stochastic partial differential equations (SPDEs) are
the basic tool for modeling systems where noise is important. SPDEs
are used for models of turbulence, pattern-formation, and the
structural development of the universe itself. It is reasonably
well-known that certain SPDEs can be manipulated to be equivalent to
(non-quantum) field theories that nevertheless exhibit deep and
important relationships with quantum field theory. In this paper we
systematically extend these ideas: We set up a functional
integral formalism and demonstrate how to extract all the one-loop
physics for an {\em arbitrary} SPDE subject to {\em arbitrary}
Gaussian noise. It is extremely important to realize that Gaussian
noise does {\em not\,} imply that the field variables undergo Gaussian
fluctuations, and that these non-quantum field theories are fully
interacting. The limitation to one-loop is not as serious as might be
supposed: Experience with quantum field theories (QFTs) has taught us
that one-loop physics is often quite adequate to give a good
description of the salient issues. The limitation to one loop does,
however, offer marked technical advantages: Because at one loop almost
any field theory can be rendered finite using zeta function
technology, we can sidestep the complications inherent in the
Martin--Siggia--Rose formalism (the SPDE analog of the BRST formalism
used in QFT) and instead focus attention on a minimalist approach that
uses only the physical fields (this ``direct approach'' is the SPDE
analog of canonical quantization using physical fields.) After setting
up the general formalism for the characteristic functional (partition
function), we show how to define the effective action to all loops,
and then focus on the one--loop effective action, and its
specialization to constant fields: the effective potential. The
physical interpretation of the effective action and effective
potential for SPDEs is addressed and we show that key features carry
over from QFT to the case of SPDEs.  An important result is that the
{\em amplitude} of the two-point function governing the noise acts as
the loop-counting parameter and is the analog of Planck's constant
$\hbar$ in this SPDE context. We derive a general expression for the
one-loop effective potential of an arbitrary SPDE subject to
translation-invariant Gaussian noise, and compare this with the
one-loop potential for QFT.

\bigskip


} 


\newcommand{\Str}{\mathop{\mathrm{Str}}}
\newcommand{\tr}{\mathop{\mathrm{tr}}}
\newcommand{\define}{\mathop{\stackrel{\mathrm def}{=}}}
\newcommand{\Tr}{\mathop{\mathrm{Tr}}}

\def\mathrm{ }
\def\d{{\mathrm{d}}}
\def\a2{a_{d/2}}
\def\implies{\Rightarrow}
\def\dirac{\gamma^\mu (\partial_\mu - A_\mu)}
\def\half{ {\scriptstyle{1\over2}} }
\def\A{ {\cal A} }
\def\PP{ {\cal P} }
\def\PPP{ {\bf P} }
\def\iff{\Leftrightarrow}
\def\J{ {\cal J} }
\def\V{ {\cal V} }

\def\Mezard{M\'ezard}

\section{Introduction}

Stochastic partial differential equations (SPDEs) are an essential
tool in modeling systems where noise is relevant~\cite{MSR}. SPDEs are
used for models of many macroscopic systems, from
turbulence~\cite{FNS,Frisch,Lvov-Procaccia}, to
pattern-formation~\cite{KPZ,MHKZ}, to the structural development of
the Universe
itself~\cite{Berera-Fang,Hochberg-Mercader,PGHL,BDGP,Dominguez-et-al}.
It is known that certain SPDEs can be studied with tools that
transform them into equivalent (stochastic) field theories which
exhibit deep and important relationships with quantum field theory
(QFT). See, for example, \cite{MSR,FNS,KPZ,MHKZ}
and~\cite{De-Dominicis-Peliti,Sun-Plischke,Frey-Tauber,Zinn-Justin}.

In this paper we set up the field theoretical ``minimalist formalism''
for SPDEs, and demonstrate how to extract the one-loop physics for an
{\em arbitrary} SPDE subject to additive Gaussian noise. It is
important to realize that Gaussian noise does {\em not\,} imply that
the field degrees of freedom undergo Gaussian fluctuations: the
combined interplay of interactions and fluctuations will appear in the
third (and higher) cumulants for the field $\phi(\vec x,t)$.  Also,
the limitation to one-loop physics is not as serious as might be
supposed: Experience with quantum field theories (QFTs) has taught us
that one-loop physics is often quite adequate to give a good
description of the salient issues
\cite{G-M+L,HMPV-tutorial,HMPV-a2,Weinberg-1,Weinberg-2,Rivers}.  In
fact, in QFT, the calculation of one-loop quantities can be augmented
by means of ``Renormalization Group improved perturbation theory'',
which contains most of the relevant features of the physics to {\em
all\,} orders in the expansion parameter~\cite{HMPV-tutorial,HMPV-a2}.
(This was called ``magical perturbation theory'' by the
authors~\cite{G-M} of reference \cite{G-M+L}.) Furthermore, at
one-loop (and higher), one can also introduce the effective action and
effective
potential~\cite{Perturbative,Non-perturbative,Coleman-Weinberg,Jackiw},
tools that allow one to determine the combined effects of interactions
and fluctuations on the ground state of the system. Defining and
calculating the one-loop effective action and effective potential is
straightforward. Interpreting the physical significance of these
quantities is more subtle.  For arbitrary SPDEs it may not even be
meaningful to define a notion of physical energy. Even when the
physical energy makes sense, dissipative effects may vitiate energy
conservation (even when noise is absent). We therefore spend some
effort in establishing that certain key features of the effective
action for QFTs carry over to SPDEs. In particular, we demonstrate
that it is still meaningful to define and calculate the effective
potential and look for its minima. The minima of the effective
potential correspond to ground states of the system, and the locations
of these minima are equal to stochastic expectation values of the
fluctuating field in the presence of noise.

While it is possible to provide an abstract non-perturbative
definition of the effective action~\cite{Non-perturbative}, in order
to proceed with explicit calculations (such as for the one-loop
effective action) one needs a perturbative procedure based on an
expansion in some small parameter. A well-known procedure of this
type, the Martin--Siggia--Rose (MSR) formalism, already exists in the
literature~\cite{MSR}.  The MSR formalism invokes additional
(unphysical) ``conjugate fields'', which are generalizations of the
fictitious fields sometimes introduced to deal with the dynamics of
diffusion. These fictitious fields permit one to extend some of the
procedures of conservative physical systems to diffusion.  For
instance, Morse and Feshbach state: ``the dodge is to consider,
$\dots$, a `mirror-image' system with negative friction, into which
the energy goes which is drained from the dissipative system''.
(See~\cite{Morse-Feshbach} page 298.) In this paper we do not make use
of the conjugate field formalism of MSR and, instead, proceed in a
direct way in which we only have physical fields, (plus possibly a
nontrivial functional Jacobian that can be rewritten in terms of ghost
fields). This approach simplifies the calculation since it halves the
number of fields one has to deal with.  These two {\em alternative}
formalisms are very similar to the situation in spontaneously broken
gauge field theories, where one can use two {\em equivalent\,}
approaches to perturbation theory, such as ``unitary gauges'' versus
``renormalizable gauges'': in one case the {\em particle content\,} is
explicit and in the other {\em renormalizability} is explicit.

After setting up the path integral formalism for the characteristic
functional (partition function), $Z[J]$, we define both the
perturbative~\cite{Perturbative} and non-perturbative effective
action~\cite{Non-perturbative}. We then focus on the one-loop
effective action and its restriction to constant (homogeneous and
stationary) fields: the effective potential~\cite{Weinberg-2,Rivers}.
An important result is that the {\em amplitude} of the noise two-point
correlation function acts as the loop-counting parameter and is the
analog of Planck's constant $\hbar$ in this SPDE context.

We conclude by deriving the formula for the one-loop effective
potential of a general SPDE subject to translation-invariant Gaussian
noise. This formula has a strong resemblance to that obtained for
ordinary QFTs and allows us to extend the use of QFT tools in the
analysis of the SPDE's effective potential.  We furthermore
demonstrate that much of the physical intuition regarding the
effective action in QFTs also carries over into SPDEs. Finally, we
offer a discussion of our results.  A number of more technical issues
are relegated to the Appendices.

\section{Stochastic partial differential equations}

\subsection{Elementary definitions}

Consider the class of stochastic partial differential equations of  
the form
\begin{equation}
D \phi(\vec x,t) = F[\phi(\vec x,t)] + \eta(\vec x,t),
\label{E:eom0}
\end{equation}
where $D$ is any linear differential operator, involving arbitrary
time and space derivatives, which does {\em not} explicitly involve
the field $\phi$. Typical examples are
\begin{eqnarray}
D &=& {\partial\over\partial t} - \nu \vec\nabla^2
\qquad \quad \hbox{Diffusion equation},
\\
D &=& {\partial^2\over\partial t^2} - \vec\nabla^2
\qquad \quad \hbox{Wave equation},
\\
D &=& {\partial\over\partial t}
\qquad \qquad \qquad  \hbox{Langevin equation}.
\end{eqnarray}
The function $F[\phi]$ is any forcing term, generally nonlinear in the
field $\phi$. Typical examples are
\begin{eqnarray}
F[\phi] &=& +{\lambda\over2} (\vec \nabla\phi)^2
\qquad \hbox{in the Kardar--Parisi--Zhang (KPZ) equation},
\\
F[\phi] &=& P[\phi]
\qquad\qquad\; \hbox{in reaction--diffusion--decay systems
($P$ is a polynomial)},
\\
F[\phi] &=& -{\delta H[\phi]\over\delta\phi}
\qquad\;\; \hbox{in ``purely dissipative'' SPDEs}.
\end{eqnarray}
The forcing term will typically not contain any time derivatives, but
this is not an essential part of the following analysis except insofar
as time derivatives may complicate some of the Jacobian functional
determinants that will be encountered below. Non-derivative terms
linear in the field can be interpreted either as decay rates or (if a
diffusion term is also present) as mass terms. They can be freely
moved between the differential operator $D$ and the forcing term
$F[\phi]$. If they are considered part of the forcing term then
\begin{eqnarray}
F[\phi] &=& -\gamma \phi
\qquad  \qquad \hbox{describes a decay term};
\\
F[\phi] &=& - \nu m^2 \phi
\qquad \;\; \hbox{describes a mass term}.
\end{eqnarray}
The function $\eta(\vec x,t)$ is a random function of its arguments
and describes the noise that we assume is present in the system.  For
the remainder of this paper, we consider field-independent additive
noise.  At this stage the nature and probability distribution of the
noise are completely arbitrary and do not need to be specified.

The noise represents our ignorance about precise details in the
dynamics of the system. It could be due, for example, to fluctuations
intrinsic to the dynamics (as in the case of Quantum Mechanics), or it
could be thought of as representing the dynamics of short-scale
degrees of freedom which have not completely decoupled from the
macroscopic dynamics ({\em e.g.}, thermal or turbulent noise), or it
could be a way of implementing ignorance of the {\em exact} initial or
boundary conditions in the system. Noise can also be a way of
summarizing the necessary truncation of the deterministic dynamics of
a many body system when we try to describe it via a finite set of
variables ({\em e.g.}, a truncated BBGKY hierarchy).

If we think of turning off the noise, we do {\em not} require that the
non-stochastic partial differential equation $D\phi = F[\phi]$ be
derivable from an action principle ({\em i.e.}, the non-stochastic
partial differential equation need not arise from a Lagrangian
formalism).  Nevertheless, once we include noise we demonstrate that
the presence of noise automatically leads to a generalized action
principle for the noisy system. It turns out that in the presence of
Gaussian noise an equation of motion {\em proportional} to the factor
$(D\phi - F[\phi])$ can always be derived by varying a well-defined
``classical'' action, and that the solutions to this equation of
motion will coincide with those of the non-stochastic equation,
provided a certain Jacobian determinant is nonsingular ({\em i.e.},
invertible).  This is explained in full detail in Appendix B.

\subsection{Some typical examples}

An example of considerable interest is the reaction--diffusion--decay
system where the SPDE is taken to be~\cite{vanKampen,Gardiner,HMPV-rdd}
\begin{equation}
\frac{\partial \phi}{\partial t}
- \nu \vec\nabla^2  \phi =
P[\phi] + \eta.
\end{equation}
This equation is used, for example, to describe the density $\phi(\vec
x,t)$ of some chemical species as a function of space and time when
the chemical is subject to both diffusion (via $\nu$) and reaction or
decay [via $P(\phi)$, a polynomial in the density field]. Expanding
out the first few terms,
\begin{equation}
P(\phi) = P_0 + P_1 \phi + P_2\phi^2 + P_3 \phi^3 +\cdots,
\end{equation}
we can identify $P_0$ with a constant (in space and time) source or
sink, $- P_1$ with the decay rate, and $P_2$ with the reaction rate
for the two-body reaction, {\em etc.}  The noise accounts for random
effects due to coupling to external sources, truncation of degrees of
freedom, averaging over microscopic effects, {\em etc.}  For $n$
species of chemical reactant the field $\phi$ is simply promoted to a
vector in configuration space $\phi(\vec x, t) \rightarrow \phi_i(\vec
x, t), \{i=1, \dots , n\}$. (The diffusion constant, $\nu$, and decay
rates then become matrices, the noise a vector, and the polynomial
$P_i(\phi_j)$ a vector-valued polynomial with tensorial coefficients.)

A second well-known example is the massive KPZ equation (equivalent to
the massive noisy Burgers equation)
\cite{KPZ,MHKZ,Sun-Plischke,Frey-Tauber,HMPV-kpz}
\begin{equation}
\frac{\partial \phi}{\partial t}
 -\nu \vec\nabla^2 \phi = -\nu m^2 \phi
+{\lambda\over2} (\vec\nabla\phi)^2 + \eta .
\end{equation}
In the fluid dynamical interpretation of the KPZ equation the fluid
velocity is taken to be $\vec v = - \vec\nabla \phi$. This model
problem leads to a form of ``turbulence'' which is known in the
literature as Burgulence~\cite{Burgulence:1,Burgulence:2}.

A third example is the enormous class of SPDEs known as ``purely
dissipative'' systems~\cite{Zinn-Justin}. Purely dissipative systems
have SPDEs of the form
\begin{equation}
{\partial \phi\over\partial t} =
- {\delta H[\phi]\over\delta\phi(\vec x)} +  \eta.
\end{equation}
These equations fall into our classification of general SPDEs as
particular types of Langevin equations with $D=\partial_t$ and with a
driving term that is a (functional) gradient $F[\phi] = -\delta
H[\phi]/\delta\phi(\vec x)$. The nomenclature ``purely dissipative''
is justified by the fact that in the {\em absence} of noise these
systems satisfy
\begin{equation}
{\partial H[\phi]\over\partial t}
= -\int \left({\delta H[\phi]\over\delta\phi(\vec x)}\right)^2 d^d \vec x 
\leq 0.
\end{equation}
Note that the reaction--diffusion--decay system can be interpreted as an
example of a purely dissipative system if we take $D=\partial_t$ and
\begin{equation}
H_{\mathrm RDD}[\phi] =
\int \left[
{\nu\over2}(\vec \nabla\phi)^2 + \int_0^{\phi(x)}  P(\tilde\phi) \;  
\d\tilde\phi
\right] \d \vec x.
\end{equation}
On the other hand, the KPZ system is {\em not} a purely dissipative
system,
\begin{eqnarray}
{\delta F_{\mathrm KPZ}[\phi(x)] \over \delta\phi(y)} &=& 
\nu m^2 \delta(x-y) 
+ \lambda \vec \nabla_x \phi(x) \cdot \vec \nabla_x \delta(x-y) 
\nonumber\\ 
&=& \nu m^2 \delta(x-y) 
- \lambda \vec \nabla_y \phi(y) \cdot 
\vec \nabla_y \delta(x-y) 
\nonumber\\ 
&\neq& {\delta F_{\mathrm KPZ}[\phi(y)] \over \delta\phi(x)}.
\end{eqnarray}
The class of purely dissipative SPDEs is a very wide one, but there
are many SPDEs that are not of purely dissipative type. We do {\em
not} want to restrict attention to purely dissipative systems in this
article and keep the discussion as general as possible.

\section{Stochastic averages, characteristic functional, Feynman  
rules}

We will focus on the stochastic partial differential equation
\begin{equation}
D\phi(\vec x,t) = F[\phi(\vec x,t)] + \eta(\vec x,t),
\label{E:stochastic1}
\end{equation}
and analyze it using functional integral techniques: Feynman diagrams,
the effective action, and the effective potential.  We develop the
field theory via the most direct route, with no conjugate fields
present.

We postpone to subsequent papers more technically involved approaches
such as the Martin--Siggia--Rose Lagrangian (with its extra unphysical
conjugate fields used for book-keeping purposes) and the hidden BRST
supersymmetry implicit in these stochastic differential
equations~\cite{MSR,De-Dominicis-Peliti,Zinn-Justin,Gozzi}.

In this section we develop the necessary tools to construct the basic
field theory and non--equilibrium statistical mechanics associated
with equation (\ref{E:stochastic1}).  We will assume {\em uniqueness}
of the solution to (\ref{E:stochastic1}), and in order to calculate
the characteristic functional, we will introduce an {\em ensemble
average} over noise realizations, and the notion of {\em delta
functionals}. Once the {\em characteristic functional} is available,
we find it useful to introduce {\em ghosts \'a la Faddeev--Popov}
before deriving the {\em Feynman rules}. We will only need to make one
assumption about the noise: that it be Gaussian, {\em i.e.,} that all
its cumulants are vanishing except the first, $\langle \eta(\vec
x,t)\rangle$, and second, $G_{\eta} = \langle \eta(\vec x,t)\eta(\vec
x',t')\rangle$.  (See {\em e.g.},~\cite{Marcienkiewicz}.)

\subsection{Step 1: Uniqueness}

Let us assume that the partial differential equation
(\ref{E:stochastic1}), plus initial conditions, is a well-posed
problem.  Thus, given a particular realization of the noise, $\eta$,
the differential equation is assumed to have a unique solution which
we designate as
\begin{equation}
\phi_{\mathrm soln}(\vec x,t|\eta).
\end{equation}
This assumption is relatively mild but does imply that the
nonlinearity is sufficiently weak so as not to drive us past a
bifurcation point.  On the other hand, it is known that noise in
concert with non--linearities can lead to the phenomenon of delayed
bifurcation in non--linear parabolic SPDEs \cite{Lythe}.  If the
partial differential equation is ill-posed, in the sense that the
solutions are not unique, additional analysis must be developed on a
case-by-case basis. A specific example of this behavior is spontaneous
symmetry breaking in QFT, which causes the naive loop expansion to
violate the convexity properties of the effective potential. This
situation must be dealt with by an improved loop
expansion~\cite{Fujimoto,Haymaker,Bender-Cooper}.

\subsection{Step 2: Ensemble average}

For any function $Q(\phi)$ of the field $\phi$ we introduce the
ensemble average (over the noise), defined by
\begin{equation}
\langle Q(\phi) \rangle \equiv
\int ({\cal D} \eta) \; \PP[\eta] \; Q(\phi_{\mathrm soln}(\vec x,  
t|\eta)),
\label{E:average1}
\end{equation}
where $\PP[\eta]$ is the probability density functional of the
noise. It is normalized to $1$, but is otherwise completely arbitrary,
that is
\begin{equation}
\int ({\cal D} \eta) \; \PP[\eta] = 1.
\end{equation}
The symbol ${\cal D}\eta$ indicates a functional integral over all
instances (or realizations) of the noise.

\subsection{Step 3: Delta functionals}

We next use a functional delta function to write the following
 identity
\begin{eqnarray}
\phi_{\mathrm soln}(\vec x,t|\eta)
&\equiv&
\int ({\cal D} \phi) \; \phi \;
\delta[\phi - \phi_{\mathrm soln}(\vec x,t|\eta)]
\nonumber\\
&=&
\int ({\cal D} \phi) \; \phi \;
\delta[D \phi- F[\phi] - \eta] \;
\sqrt{ {\cal J} {\cal J}^\dagger },
\end{eqnarray}
where we have performed a change of variables and introduced the
Jacobian functional determinant, defined by
\begin{equation}
{\cal J} \equiv \det\left( D - {\delta F\over\delta\phi} \right),
\end{equation}
and its adjoint
\begin{equation}
{\cal J}^\dagger \equiv
\det\left( D^\dagger - {\delta F\over\delta\phi}^\dagger \right).
\end{equation}
The above is just the functional analogue of a standard delta function
result: If $f(x) =0$ has a unique solution at $x=x_0$, then
\begin{equation}
x_0 =
\int \d x \; x \; \delta(x-x_0) =
\int \d x \; x \; \delta(f(x)) \; |f'(x)| =
\int \d x \; x \; \delta(f(x)) \; \sqrt{f'(x) [f'(x)]^*}.
\end{equation}
The delta function forces one to pick up only one contribution from
the solution of the equation $f(x)=0$, and the derivative is there to
provide the correct measure to the integral.  In the functional case
the derivative becomes a determinant. It is in fact the Jacobian
determinant associated with the change of variables from $\phi$ to
$D\phi-F[\phi]$. It is now easy to see that one also has the identity
\begin{equation}
Q(\phi_{\mathrm soln}(\vec x,t|\eta)) \equiv
\int ({\cal D} \phi) \; Q(\phi) \;
\delta\left( D \phi- F[\phi] - \eta\right) \;
\sqrt{{\cal J} {\cal J}^\dagger }.
\end{equation}
Furthermore, the ensemble average over the noise, equation
(\ref{E:average1}), becomes
\begin{equation}
\langle Q(\phi) \rangle =
\int ({\cal D} \eta) \; ({\cal D} \phi) \; \PP[\eta]
\; Q(\phi) \;
\delta\left( D \phi- F[\phi] - \eta\right) \;
\sqrt{{\cal J} {\cal J}^\dagger }.
\end{equation}
The noise integral is easy to perform, with the result that for
arbitrary stochastic averages one has
\begin{equation}
\langle Q(\phi) \rangle =
\int  ({\cal D} \phi)\; \PP[D\phi-F[\phi]] \; Q(\phi) \;
\sqrt{{\cal J} {\cal J}^\dagger }.
\end{equation}
We see from this equation that the effect of the noise only appears in
the stochastic average through its probability distribution
$\PP[D\phi-F[\phi]]$. It is worthwhile to point out that the main
difference, at this stage of the formalism, between the present
``minimal'' approach and that of MSR lies in the way the delta
functional is handled. In MSR, instead of integrating directly over
the noise, as is done here, the delta functional is replaced by its
functional Fourier integral representation.  This is the step wherein
the conjugate field enters.  If this latter route is taken, the noise
integration can be performed exactly only for Gaussian noise. In the
minimal formalism by contrast, the integration over the noise can be
done exactly for arbitrary noise. It thus lends itself immediately for
handling non-Gaussian systems: For general noise distributions we can
explicitly write down the probability distribution for the fields as
\begin{equation}
\PPP[\phi] = \PP[D\phi-F[\phi]] \; \sqrt{{\cal J} {\cal J}^\dagger }.
\end{equation}
We will not further explore the possibility of arbitrary noise in this
paper, since Gaussian noise (which manifestly does {\em not} imply
Gaussian fluctuations of the fields) is already sufficiently general
to be of great practical interest.

The presence of the functional determinant is essential: it must be
kept to ensure proper counting of the solutions to the original
stochastic differential equation.  In QFT this functional determinant
is known as the Faddeev--Popov determinant and is essential in
maintaining unitarity~\cite{Weinberg-1,Rivers}, {\em i.e.},
conservation of probability. In some particular cases the functional
determinant is field-independent, and it is safe to neglect it. We
discuss this more fully in Appendix A and in the companion
papers~\cite{HMPV-rdd,HMPV-kpz}, but for the sake of generality we
will carry these determinants along (with little extra cost) for the
rest of this paper.

\subsection{Step 4: Characteristic functional (Partition function)}

A particularly useful quantity is the generating functional, or
characteristic functional (partition function), defined by taking
\begin{equation}
Q(\phi) = \exp\left( \int
 \d^d\vec x \; \d t
\; J(\vec x,t) \; \phi(\vec x,t) \;
\right)
\end{equation}
in equation (\ref{E:average1}). We define it as follows
\begin{eqnarray}
Z[J] &\stackrel{\mathrm def}{=}&
\left\langle
\exp \left( \int \d x \; J(x) \; \phi(x)  \right)
\right\rangle
\\
&=& \int ({\cal D} \phi)\; \PPP[\phi]
         \exp\left( \int \d x \; J \; \phi \; \right) \;
\\
&=& \int ({\cal D} \phi)\; \PP[D \phi- F[\phi]] \;
         \exp\left( \int \d x \; J \; \phi \; \right) \;
         \sqrt{{\cal J} {\cal J}^\dagger },
\label{E:characteristic0}
\end{eqnarray}
with an obvious condensation of notation, $\d x = \d^d\vec x \; \d t$.
When there is no risk of confusion we will suppress the $\d x$
completely.  This key result will enable us to calculate the effective
action and the effective potential in a direct way.

\subsection{Step 5: Gaussian noise}

We will now make some assumptions about the noise: We assume it to be
Gaussian. Without loss of generality we can take the noise to have
zero mean, since if the mean is nonzero we can always redefine the
forcing term $F[\phi]$ to make the noise have zero mean.  We therefore
take the noise to be Gaussian of zero mean, so that the only non-zero
cumulant is the second order one.  We do not need to make any more
specific assumptions about the functional realization of the noise:
the noise might (for instance) be white, power-law, colored, pink,
$1/f$-noise, or shot noise, and our considerations below apply to all
of these cases. As long as the noise is Gaussian its probability
distribution can be written as
\begin{equation}
\PP[\eta] = {1\over\sqrt{\det(2\pi G_\eta)}}
\exp\left( -\half
\int \int \d x \; \d y \; \eta(x) \;  G_\eta^{-1}(x,y) \; \eta(y)
\right).
\end{equation}
The characteristic functional (partition function) is thus seen from
equation (\ref{E:characteristic0}) to be
\begin{eqnarray}
Z[J] &=& {1\over\sqrt{\det(2\pi G_\eta)}}
         \int ({\cal D} \phi)\;
\sqrt{{\cal J} {\cal J}^\dagger }
         \exp\left( \int J \phi \right) \;
         \exp\left( -\half
                   \int \int (D\phi-F[\phi])  G_\eta^{-1} (D\phi-F[\phi])
             \right).
\label{E:characteristic1}
\end{eqnarray}
This characteristic functional (partition function) contains all the
physics of the model since it allows for the calculation of averages,
correlation functions, thermodynamic variables, {\em etc.} Note that
the noise has been completely eliminated and survives only through the
explicit appearance of its two-point correlation function in the
above.  Since the characteristic functional is now given as a path
integral over the physical field, all the standard machinery of
statistical field theory (and quantum field theory) can be brought to
bear~\cite{Textbooks}. See, for
example~\cite{Zinn-Justin,Weinberg-1,Weinberg-2,Rivers},
and~\cite{Bjorken-Drell-1}---\cite{Le-Bellac}.

This formula for the partition function demonstrates that (modulo
Jacobian determinants) all of the physics of any stochastic
differential equation can be extracted from a functional integral
based on the ``classical action''
\begin{equation}
{\cal S}_{\mathrm classical} = \frac{1}{2}
  \int \int (D\phi-F[\phi])  G_\eta^{-1} (D\phi-F[\phi])
\; .
\end{equation}
This ``classical action'' is a generalization of the Onsager--Machlup
action~\cite{Onsager-Machlup}. The Onsager--Machlup paper dealt with
stochastic differential equations rather than partial differential
equations (mechanics rather than field theory), and was limited to
noise that was temporally white.  As their formalism was developed
with the notions of linear response theory in mind, Onsager and
Machlup assumed the ``forcing term'' $F[\phi]$ to be linear, so that
{\em both} the noise and the field fluctuations were Gaussian. In our
formalism all these assumptions can be relaxed: the forcing term can
be nonlinear and in general the field fluctuations will not be
Gaussian even if the noise is Gaussian.

\subsection{Step 6: Faddeev--Popov ghosts}

We mentioned previously that the Jacobian determinant is often (not
always) field-independent.  This is a consequence of the causal
structure of the theory as embodied in the fact that we are only
interested in {\em retarded} Green functions.  The situation here is
in marked contrast with that in QFT where the relativistic nature of the
theory forces the use of {\em Feynman} Green functions ($+i\epsilon$
prescription).  As we explain in Appendix A, this change radically
alters the behavior of the functional determinant.

In order to avoid too many special cases, and to have a formalism that
can handle both constant and field-dependent Jacobian factors, we
exponentiate the determinant via the introduction of a pair  of
Faddeev--Popov ghost fields~\cite{Zinn-Justin,Weinberg-1},
\begin{equation}
\label{E:ghost}
{\cal J} \equiv
\det\left( D - {\delta F\over\delta\phi} \right)
=
{1\over\det(2\pi I)}
\int ({\cal D} [g^\dagger,g]\,) \;
\exp\left(
-\half\int  g^\dagger
\left[ D - {\delta F\over\delta\phi} \right]
g
\right),
\end{equation}
where $I$ is the identity operator on spacetime.  The $g$ field is a
so-called (complex) scalar ghost field. It is a field of
anti-commuting complex variables and behaves in a manner similar to an
ordinary scalar field except that there is an extra minus sign for
each ghost loop. We also need to use the {\em conjugate} ghost field
$g^\dagger$ to handle the determinant.

We should point out that if the operator $D - (\delta F/\delta\phi)$
is self-adjoint then $\cal J = {\cal J}^\dagger$. In this case $\sqrt{
{\cal J} {\cal J}^\dagger}$ reduces to $|{\cal J}| = |{\cal
J}^\dagger|$.  In QFTs the relevant operators occurring in the
Jacobian determinants are obtained from second functional derivatives
of the action and are automatically self-adjoint. In contrast, for
SPDEs there is no guarantee that $D - (\delta F/\delta\phi)$ be
self-adjoint, and in fact for the examples previously discussed (KPZ,
reaction--diffusion--decay, and purely dissipative) this operator is
not self adjoint. Instead we rely on the much weaker property that the
operator $D - (\delta F/\delta\phi)$ is {\em real} in order to write
$\sqrt{ {\cal J} {\cal J}^\dagger} = \sqrt{{\cal J} {\cal J}^*} =
|{\cal J}|$. In all cases we are interested in the relevant operators
are not only real, but positive, so that the absolute value symbol can
be ignored and the characteristic functional equation
(\ref{E:characteristic1}) is given by
\begin{eqnarray}
Z[J] &=&
{1\over\sqrt{\det[(2\pi)^3 G_\eta]}}
\int ({\cal D} \phi)\;  ({\cal D} g) \;
({\cal D} g^\dagger) \;
         \exp\left( -\half
                   \int \int (D\phi-F[\phi])  G_\eta^{-1} (D\phi-F[\phi])
             \right) \;
\nonumber\\
&&\qquad
\exp\left( 
- \half \int g^\dagger  \left[ D - {\delta F\over\delta\phi} \right] g
\right)
\;         \exp\left( \int J \phi \right).
\label{E:characteristic2}
\end{eqnarray}
This procedure trades off the functional determinants for two extra
functional integrals.  The advantage of this procedure becomes clear
when one develops the perturbation theory.  (This Faddeev--Popov trick
for exponentiating the Jacobian determinant is also essential in
finding the hidden BRST supersymmetry.)

It must be noted that neither pair of ghost field variables couple to
an external source. This means they can only appear in internal lines
in Feynman diagrams, a fact that will be used later on when we discuss
loops and loop-counting.

In Appendix A we take a closer look at the Jacobian functional
determinant, its causal structure, and its specific form for local
driving forces. In the latter part of this Appendix we make use of the
perturbation theory based on Feynman diagrams to evaluate this
functional determinant from another perspective.

\subsection{Step 7: Feynman rules}

With the partition function in the form given above, (with two
independent ghosts), it is now easy to develop a formal Feynman
diagram expansion. We wish to treat the driving term $F[\phi]$ as the
perturbation and expand around the free-field theory defined by
setting $F=0$.  With this convention the free action is, explicitly,
\begin{equation}
{\cal S}_{\mathrm free} \equiv
\int \int \left\{ \half \; [D\phi] \;  G_\eta^{-1} \; [D\phi] \right\} 
\; \d x \; \d y +
\int \left\{ \half \; g^\dagger \; D \; g \right\} dx.
\end{equation}
There are two particle propagators in this free action, one for the
$\phi$ field, and two for the ghost fields. Formally
\begin{eqnarray}
G_{\phi\phi} &=& [ D^\dagger \;  G_\eta^{-1} \; D ]^{-1}
                   =  [ D^{-1} \; G_\eta (D^\dagger)^{-1} ];
\\
G_{g^\dagger g} &=& [ D ]^{-1}.
\end{eqnarray}
Here $D^\dagger$ is the adjoint operator of $D$, defined by partial
integration. (For instance if $D=\partial_t -\nu \vec \nabla^2$ then
$D^\dagger = -\partial_t - \nu \vec \nabla^2$.)  In the interests of
generality, we reiterate the fact that we have not assumed translation
invariance for the noise, (although the noise is now Gaussian.) The
momentum-frequency representation for the propagators is
\begin{eqnarray}
G_{\phi\phi}(\vec k_1,\omega_1;\vec k_2,\omega_2) &=&
{ G_\eta(\vec k_1,\omega_1;\vec k_2,\omega_2)
\over D^\dagger(\vec k_1,\omega_1) \; D(\vec k_2,\omega_2)};
\\
G_{g^\dagger g}(\vec k,\omega) &=& {1\over D(\vec k,\omega)};
\end{eqnarray}
The Feynman vertices come from the interaction piece of the action,
which in this convention is
\begin{equation}
{\cal S}_{\mathrm interaction}  =
\int \int
\left\{
-[D\phi] \;  G_\eta^{-1} \; F[\phi] +
\half F[\phi] \;  G_\eta^{-1} \;F[\phi]
\right\} \d x \,\d y
- \int \left\{ \half \; 
g^\dagger  \; {\delta F\over\delta\phi} \; g\right\} \d x.
\end{equation}
The nature of the vertices (obtained by functional differentiation
with respect to the fields present in the theory) depends on the
structure of the forcing term $F[\phi]$. In the meantime we formally
assert
\begin{eqnarray}
&\phi-F[\phi]\hbox{ vertex:}& \qquad
-{D(\vec k_1,\omega_1) \; \phi \; F[\phi]\over
 G_\eta(\vec k_1,\omega_1;\vec k_2,\omega_2)};
\\
&F[\phi]-F[\phi] \hbox{ vertex:}&  \qquad
+ \half \; { F[\phi] \; F[\phi] \over
G_\eta(\vec k_1,\omega_1;\vec k_2,\omega_2)};
\\
&\hbox{ghost vertex:}& \qquad
- \half \; g^\dagger\; {\delta F[\phi]\over\delta\phi}\; g.
\end{eqnarray}
Note that to turn these schematic Feynman rules into practical
computational tools we will need to assume that $F[\phi]$ is some
specific local functional of the field $\phi$. (Typically a polynomial
or polynomial with derivatives.) When we define the effective action
we will again see that the formalism can be successfully developed
even for non-translation-invariant noise, and this derivation of the
Feynman rules matches the generality of the definition of the
effective action. This concludes, for now, the most general aspects of
the discussion.

When it comes to actual calculations in specific models, the majority
of these models have noise that is not only Gaussian but is also
translation invariant. In the interests of simplicity, we now
(finally, and only for the rest of this particular section) indicate
the effects of assuming translation invariance for the noise.  This
lets us take simple Fourier transforms in the difference variable
$x-y$ (more precisely: $\vec x-\vec y$ and $t_x-t_y$) to see that in
momentum-frequency space
\begin{eqnarray}
G_{\phi\phi}(\vec k,\omega) &=&
{ G_\eta(\vec k,\omega)\over D^\dagger(\vec k,\omega) \; D(\vec  
k,\omega)} =
{ G_\eta(\vec k,\omega)\over D(-\vec k,-\omega) \; D(\vec k,\omega)};
\\
G_{g^\dagger g}(\vec k,\omega) &=& {1\over D(\vec k,\omega)}.
\end{eqnarray}
The Feynman diagram vertices are now ($\phi$ and $g$ are here
understood to be Fourier transformed)
\begin{eqnarray}
&\phi-F[\phi]\hbox{ vertex:}& \qquad
-{D(\vec k,\omega) \;\phi  \; F[\phi]\over  G_\eta(\vec k,\omega)};
\\
&F[\phi]-F[\phi] \hbox{ vertex:}&  \qquad
+\half{ F[\phi] \; F[\phi] \over G_\eta(\vec k,\omega)};
\\
&\hbox{ghost vertex:}& \qquad
- \half \; g^\dagger \; {\delta F[\phi]\over\delta\phi} \; g.
\end{eqnarray}
As always, there is a certain amount of freedom in writing down the
Feynman rules. It is always possible to take part of the quadratic
piece in the total action and move it from the free action to the
interaction term or vice versa. We have already seen that a linear
term ({\em e.g.}, $\nu m^2 \phi$) in the forcing function $F[\phi]$
can with equal facility be reassigned to the differential operator $D$
via the scheme $D \to D - \nu m^2$, $F[\phi]\to F[\phi] - \nu m^2
\phi$. This procedure can always be used to completely eliminate any
linear term in $F[\phi]$.  Similar but more complicated behaviour
occurs if the forcing function contains both constant and quadratic
pieces ({\em e.g.}, $a+b\phi^2$).  With the conventions given above
the interaction term contains (at least) a ``cosmological constant'',
$\half a^2 \int \int G_\eta^{-1}(x,y) dx dy$, a quadratic piece, $ab
\int \int G_\eta^{-1}(x,y) \phi^2(y) dx dy$, and a $\phi^4$
interaction. The quadratic piece could be moved into the free action
at the cost of making the expression for the free propagator a little
more complicated. This freedom in writing down the Feynman rules does
not imply any ambiguity in the physical results: Moving quadratic
pieces around from the interaction term to the free action will modify
the Feynman rules but will not affect any physical quantities.

\section{Effective Action: Loop expansion}

In order to set up the formalism for the effective action, and its
loop expansion, it is useful to first separate the two-point function
for the noise into a {\em shape}, $g_2(x,y)$, and a constant {\em
amplitude}, $\A$, via the correspondence
\begin{equation}
G_\eta(x,y) \define \A \; g_2(x,y).
\end{equation}
For the case of Gaussian white noise this is automatically satisfied
by definition: $G_\eta(x,y) \to \A \; \delta(x-y)$. For more general
Gaussian noises (which describe, for instance, the effects of small
scale degrees of freedom not fully decoupled from the physics of
$\phi$, such as in the case of a heat bath into which $\phi$ has been
immersed) this form of the two-point function allows the
interpretation of the noise intensity $\A$ as a characterization of
the bath-system coupling. This will become more evident when we
compare the ``effective potential'' in noisy environments, equation
(\ref{E:general}) below, with the same object for zero temperature
quantum field theory, equation (\ref{E:general1}).  The normalization
of the shape function $g_2(x,y)$ is essentially arbitrary, and any
convenient normalization will suffice.

Another advantage of singling out the intensity parameter $\A$ is that
it is the loop-counting parameter for this formulation of SPDEs.  To
see this, one starts by writing the characteristic functional (with
external sources rescaled for convenience) as
\begin{eqnarray}
Z[J] &=&
{1\over\sqrt{\det[(2\pi)^3 G_\eta]}}
\int ({\cal D} \phi)\;  ({\cal D} g) \;
({\cal D} g^\dagger) \;
         \exp\left( -\half
             \int \int { (D\phi-F[\phi])  g_2^{-1} (D\phi-F[\phi])  
\over\A }
             \right) \;
\nonumber\\
&&\qquad
\exp\left(
-\half\int g^\dagger \left[ D - {\delta F\over\delta\phi} \right] g
\right)
\;
\exp\left( {\int J \; \phi\over \A} \right).
\label{E:characteristic3}
\end{eqnarray}
The generating function (Helmholtz free energy in statistical field
theory) for connected correlation functions is defined by
\begin{equation}
W[J] = + \A \left\{ \ln Z[J] - \ln Z[0] \right\}.
\end{equation}
The effective action (Gibbs free energy in statistical field theory)
is then defined {\em non-perturbatively} in terms of $W[J]$ by taking
its Legendre transform~\cite{Weinberg-2,Non-perturbative}
\begin{equation}
\Gamma[\phi;\phi_0] = - W[J] + \int \phi \; J;
\qquad\qquad
{\delta W[J]\over \delta J} = \phi;
\qquad\qquad
{\delta \Gamma[\phi;\phi_0]\over \delta \phi} = J.
\end{equation}
Here $\phi_0$ is some suitable background (mean) field, which is taken
to be the stochastic expectation value of $\phi$ in the absence of
external sources, $J=0$. It is often but not always zero, and we
retain it for generality.

The previous equation defines the non-perturbative effective
action. For any specific example the previous equation is not very
useful, and we often have to restrict ourselves to a perturbative
calculation of the effective action, after singling out an expansion
parameter.  One can always develop a Feynman diagram expansion
provided that the classical action can be separated into a ``quadratic
piece'' and an ``interacting term'', as we have already done.

In the loop expansion the sum of all connected diagrams coupled to
external sources $J(x)$ is exactly $W[J]$ as defined above, and the
effective action $\Gamma[\phi;\phi_0]$ corresponds to all (amputated)
one-particle irreducible graphs (1PI), that is, Feynman diagrams that
cannot be made disconnected by cutting only one propagator.

In the following argument, we will be considering diagrams
contributing to the effective action. Recall that ghosts can only
appear as internal lines, since they are not coupled to external
sources.

To see the role of the amplitude $\A$ as loop-counting parameter, note
that each field propagator is proportional to $\A$ while each ghost
propagator is independent of $\A$. The vertices that do not include
ghosts are proportional to $\A^{-1}$, while ghost vertices are
independent of $\A$. Thus each Feynman diagram contributing to the
effective action is proportional to $A^{I_\phi-V_\phi}$, where
$I_\phi$ is the number of non-ghost propagators, and $V_\phi$ is the
number of non-ghost vertices.  But each ghost vertex is attached to
exactly two ghost propagators (except for tadpole ghost loops), and
each ghost propagator is attached to exactly two ghost vertices
(except for tadpole ghost loops). In the case of tadpole ghost loops,
exactly one propagator is attached to exactly one ghost vertex. This
implies that if one assigns a factor $\A$ to each ghost propagator and
a factor of $\A^{-1}$ to each ghost loop then one will not change the
total number of factors of $\A$ assigned to the Feynman diagram. Thus
the Feynman diagrams are proportional to $\A^{I-V}$ where $I$ is the
total number of (internal) propagators in the Feynman diagram and $V$
is the total number of vertices, now including ghosts.

It is the result of a standard topological theorem that for any graph
(not just any Feynman diagram) $ I - V = L - 1 $, where $L$ is the
number of loops~\cite{Weinberg-1,Rivers,Itzykson-Zuber}.  It is then
easy to see that field theories based on SPDEs exhibit exactly the
same loop-counting properties as QFTs except that the loop-counting
parameter is now the {\em amplitude} of the noise two-point function
(instead of Planck's constant $\hbar$).  The only subtle part of the
argument has been in dealing with the Faddeev--Popov ghosts, and it is
important to realize that this argument is completely independent of
the details of the differential operator $D$ and the forcing term
$F[\phi]$. When it comes to calculating the diagrams contributing to
the effective action, the extra explicit factor of $\A$ inserted in
the definition of $W[J]$ above guarantees that the 1PI graphs
contribute to $\Gamma[\phi;\phi_0]$ with a weight that is exactly
$\A^L$.  This demonstrates that $\A$ is a {\em bona fide} expansion
parameter.

At this point, it becomes natural to make a comparison with the MSR
(Martin--Siggia--Rose) formalism for the calculation of the effective
action in Stochastic Field Theories, where one introduces a field {\em
conjugate} to $\phi$. Historically, this conjugate field first arose
in setting up a variational approach to the diffusion equation, ({\em
cf.} Morse and Feshbach {\em loc. cit.}). The following remarks will
help one to understand the differences and the complementarity of our
approach to the MSR approach; the bottom line is related to technical
issues associated to proving all-orders
renormalizability~\cite{MSR,De-Dominicis-Peliti,Zinn-Justin}.  The
direct approach developed in this paper is akin to the ghost-free
axial gauge of QCD or the so-called unitary gauge in the standard
model of particle physics: This is a formalism well-adapted to
isolating the physical degrees of freedom, at least perturbatively,
but is not well-adapted to proving the all-orders renormalizability of
the theory. (Proving one-loop renormalizability for specific theories
is not too difficult, and we will address this issue in a pair of
companion papers~\cite{HMPV-rdd,HMPV-kpz}.)

In analogy with the situation in QFT, one has three possible
responses to this state of affairs:
\begin{enumerate}
\item 
Use the MSR formalism for all calculations. This is comparable to
  using BRST--invariant versions of the standard model of particle
  physics to calculate scattering cross-sections and decay rates.
  (That is, overkill.)
\item 
One could appeal to the fact that the SPDEs considered in this paper
  are hardly likely to be thought of as fundamental theories in the
  particle physics sense; These SPDEs are much more like ``effective
  field theories'', in that the noise and fluctuations in real
  physical systems are manifestations of our lack of knowledge of the
  short-distance physics. Viewed as effective theories,
  renormalizability is no longer the main guiding light it was once
  thought to be~\cite{Weinberg-2}.
\item 
At a very {\em practical} level one can choose to be guided by
  experience with quantum field theories. It is well known that
  one-loop physics is often sufficient for extracting most of the
  physical information from a system. Calculations beyond one-loop,
  while certainly important at a fundamental level, are often more
  than is really needed. One of the great technical simplifications of
  one-loop physics is that, via zeta function technology, essentially
  any field theory can be regularized at one-loop without excessive
  complications~\cite{Dowker,DeWitt,BVW}.
\end{enumerate}

For these reasons we will now restrict our attention to a one-loop
calculation (apart from the discussion of Feynman diagrams and the
loop expansion everything up to this point has been valid
non-perturbatively, while those discussions were still valid to
all orders in perturbation theory). In the next section we calculate
the one-loop effective action.

\section{Effective Action: One Loop}

It is well known that the effective action for a field theory can be
obtained by performing a Legendre transform on the logarithm of the
characteristic functional (partition function). Writing
\begin{equation}
Z[J] = \int {\cal D}\phi
\exp\left({ - {\cal S}[\phi] + \int J\; \phi \over a}\right),
\label{E:characteristic4}
\end{equation}
where {\em a} is the parameter characterizing the fluctuations, one
gets for the one-loop effective action (first order in {\em a})
\begin{equation}
\Gamma[\phi;\phi_0] = {\cal S}[\phi] - {\cal S}[\phi_0] +
\half a \left\{ \ln \det ({\cal S}_2[\phi]) -
                \ln \det ({\cal S}_2[\phi_0]) \right\}
+ O(a^2).
\label{E:effective}
\end{equation}
Here ${\cal S}_2 \define \delta^2 {\cal S}/\delta \phi(x) \delta
\phi(y)$ is the matrix of second order functional derivatives of the
action ${\cal S}[\phi]$ (often called the Jacobi field operator).  For
QFT the loop-counting parameter $a$ is Planck's constant $\hbar$, and
${\cal S}_2$ is a second-order partial differential operator that
depends on the field $\phi$ via some potential-like term. The
determinants of partial differential operators can be defined and
calculated by a variety of techniques.  The notation ${\cal
S}[\phi_0]$ is actually shorthand for ${\cal
S}[\langle\phi[J=0]\rangle]$, and for a symmetric ground state
($\langle\phi[J=0]\rangle=0$) one often has ${\cal S}[0]=0$. These
terms contribute a constant offset to the effective action. In QFT
these terms are interpreted as a field-independent contribution to the
vacuum energy and are traditionally ignored, although in the context
of cosmology, they contribute (sometimes catastrophically) to the
cosmological constant. In the interest of generality we will make them
explicit. When we consider field theories based on SPDEs, the
loop-counting parameter {\em a} becomes $\A$, which we singled out as
the amplitude for the noise, and the bare action in equation
(\ref{E:characteristic3}) is replaced by equations
(\ref{E:characteristic1}) and (\ref{E:characteristic2})
\begin{eqnarray}
{\cal S}[\phi]
&\to&
\half \int \int
\left\{ (D\phi-F[\phi]) g_2^{-1} (D\phi-F[\phi]) \right\}
\d^d \vec x \; \d t\; \d^d \vec y \; \d t'
- \half\A \; \left(
\ln {\cal J} +
\ln {\cal J}^\dagger
\right)
\\
&=& {\cal S}_{\mathrm classical}[\phi]
- \half\A \; \left(
\ln {\cal J} +
\ln {\cal J}^\dagger
\right),
\label{E:bareaction}
\end{eqnarray}
where on the second line we have denoted by ${\cal S}_{\mathrm
  classical}[\phi]$ the double integral in the previous line. This is
  the quantity that we have previously defined as the nonlinear
  generalization of the Onsager--Machlup action to arbitrary Gaussian
  noise~\cite{Onsager-Machlup}.

The noise at this stage is Gaussian, and does not need to be
translation invariant. We have explicitly kept the Jacobian functional
determinant.  Inserting equation (\ref{E:bareaction}) into the formula
for the one-loop effective action [equation (\ref{E:effective})], we
obtain the following general result (applicable to any SPDE) 
\begin{eqnarray}
\Gamma[\phi;\phi_0] &=&
{\cal S}_{\mathrm classical}[\phi] -
{\cal S}_{\mathrm classical}[\phi_0]
\nonumber\\
&&\qquad +
{\cal A} \left\{
\half \ln \det ({\cal S}_2[\phi]) - \half \ln \det ({\cal S}_2[\phi_0]) 
-\half\ln {\cal J}[\phi]-\half\ln {\cal J}^\dagger[\phi]
+\half\ln {\cal J}[\phi_0] +\half\ln {\cal J}^\dagger[\phi_0]
\right\}
\nonumber\\
&&\qquad
+ O({\cal A}^2).
\end{eqnarray}
To make this more explicit, the fluctuation operator ${\cal
S}_2(\phi)$ ({\em aka} Jacobi field operator) is
\begin{equation}
{\cal S}_2[\phi] =
\left( D^{\leftarrow} - {\delta F\over \delta\phi}^{\leftarrow} \right)
g_2^{-1}
\left(  D - {\delta F\over \delta\phi} \right) -
(D\phi-F[\phi])  g_2^{-1} {\delta^2 F\over\delta\phi\;\delta\phi}.
\end{equation}
Here the $\leftarrow$ indicates that these operators should be thought
of as acting to the left. Also $g_2^{-1}(x,y)$ is to be understood as
a ``matrix'' with implicit sums over the indices $x,y$ ({\em i.e.},
integrations over the variables.) Note that if $F[\phi]$ contains
derivatives of $\phi$ then $\delta F/\delta\phi$ will be a
differential operator.  Performing an integration by parts, this can
be converted to a statement about the adjoint operator acting to the
right, {\em i.e.,} we can re-write ${\cal S}_2[\phi]$ as
\begin{equation}
{\cal S}_2[\phi] =
\left( D^\dagger - {\delta F\over \delta\phi}^\dagger \right)
g_2^{-1}
\left(  D - {\delta F\over \delta\phi} \right) -
(D \phi-F[\phi])  g_2^{-1} {\delta^2 F\over\delta\phi\;\delta\phi}.
\end{equation}
Putting all this together gives the following one-loop result for the
effective action
\begin{eqnarray}
\label{E:Gamma-1}
\Gamma[\phi;\phi_0] &=&
\half \int \int \d^d \vec x \; \d t \; \d^d \vec y \; \d t'
\left\{ (D \phi-F[\phi]) g_2^{-1}  (D \phi-F[\phi]) \right\}
- \half\A \; \left(
\ln {\cal J} +
\ln {\cal J}^\dagger
\right)
\nonumber\\
&&+ \half \A \ln \det
\left[
\left( D^\dagger  - {\delta F\over \delta\phi}^\dagger \right)
 g_2^{-1}
\left(  D - {\delta F\over \delta\phi} \right) -
(D \phi-F[\phi])  g_2^{-1}  {\delta^2 F\over\delta\phi\;\delta\phi}
\right]
\nonumber\\
&&- \left( \phi \to \phi_0 \right)
+ O(\A^2).
\end{eqnarray}
Grouping together the terms proportional to $\A$, and using the
representation of the functional determinant, enables us to rewrite
the above in the alternative form
\begin{eqnarray}
\label{E:Gamma-2}
\Gamma[\phi;\phi_0] &=&
\half\int \int \d^d \vec x\; \d t \; \d^d \vec y \; \d t'
\left\{ (D \phi-F[\phi])  g_2^{-1} (D \phi-F[\phi]) \right\}
\nonumber\\
&+& \half \A \ln \det
\left[ I -
\left\{
\left( D^\dagger - {\delta F\over \delta\phi}^\dagger \right)^{-1}
 g_2
\left(  D - {\delta F\over \delta\phi} \right)^{-1}
\left(
(D \phi-F[\phi]) g_2^{-1}  {\delta^2 F\over\delta\phi\;\delta\phi}
\right)
\right\}
\right]
\nonumber\\
&&- \left( \phi \to \phi_0 \right)
+ O(\A^2).
\end{eqnarray}
This expression for the one-loop effective action is instructive.  It
is made up of two contributions whose origin and physics are quite
different. On the one hand, the first term (the generalized
Onsager-Machlup term) gives a contribution whose form is directly
related to both the noise {\em shape factor} and the non-noisy part of
the equation of motion, including non-linearities.  On the other hand,
the log-determinant term is proportional to the noise {\em amplitude}
(which we have seen is the loop expansion parameter) and its specific
form depends {\em also} on the structures of $D$ and $F[\phi]$, as
well as on properties of the noise shape function.  Therefore, noise
plays a central role in the physics of the SPDE and, as will be
discussed below, particularly in the nature of the ground state of the
stochastic system described by equation (\ref{E:eom0}).

\section{Effective Potential: One Loop}

We now concentrate on field configurations that are homogeneous and
static. For such field configurations the effective action reduces to
a quantity known as the ``effective potential''. In this section we
will {\em calculate} the effective potential, deferring the discussion
of its physical {\em interpretation} (in the context of SPDEs) to the
next section.

The effective potential is defined as
\begin{equation}
{\cal V}[\phi;\phi_0] = {\Gamma[\phi;\phi_0] \over \Omega},
\end{equation}
with $\phi$ a homogeneous and static field configuration, and
$\Omega$ the volume of spacetime. The effective potential at one loop
is given by
\begin{eqnarray}
{\cal V}[\phi;\phi_0] &=&
\half F^2[\phi] \left\{ \int \d^d \vec x \; \d t \; g_2^{-1}
\right\}
- \half {\A\over\Omega} \ln \det
\left( D - {\delta F\over \delta\phi} \right)
- \half {\A\over\Omega} \ln \det
\left( D^\dagger - {\delta F\over \delta\phi}^\dagger \right)
\nonumber\\
&&+ \half {\A\over\Omega} \ln \det
\left[
\left( D^\dagger  - {\delta F\over \delta\phi}^\dagger \right)
 g_2^{-1}
\left(  D - {\delta F\over \delta\phi} \right) +
F[\phi] \left\{ \int \d^d \vec x \; \d t \; g_2^{-1} \right\}
{{\delta^2 F}\over{\delta\phi\;\delta\phi}}
\right]
\nonumber\\
&&- \left( \phi \to \phi_0 \right)
+ O(\A^2).
\label{E:ep0}
\end{eqnarray}
In order to turn this into a more tractable expression it is useful to
introduce a frequency-momentum representation. First notice that
\begin{eqnarray}\label{noisenorm}
\int \d^d \vec x \; \d t  \;  g_2^{-1}(\vec x,t)
&=&
\int  {\d^d \vec k \; \d\omega \over(2\pi)^{d+1}} \; \d^d \vec x \; \d t
\; \tilde g_2^{-1}(\vec k,\omega) \; \exp[-i(\omega t - \vec k \cdot
\vec x)]
\nonumber\\
&=& \tilde g_2^{-1}(\vec k=\vec 0,\omega=0).
\end{eqnarray}
(It is clear from the formula for the one-loop effective potential
equation (\ref{E:ep0}) that the above integral has to be finite, or
rendered finite by appropriate renormalizations of the noise
correlation function and the parameters it contains.)

We next make use of the following identity valid for a translation
invariant operator $X$
\begin{eqnarray}
\ln \det X
&=&
\int \d^d \vec x \; \d t \;
\int \d^d \vec k_1 \; \d \omega_1 \;
\int \d^d \vec k_2 \; \d \omega_2 \;
\langle \vec x,t | \vec k_1,\omega_1 \rangle \;
\ln X(\vec k_1, \omega_1) \; \delta^d(\vec k_1,\vec k_2) \;
\delta(\omega_1,\omega_2) \;
\langle \vec k_2, \omega_2 |\vec x,t \rangle
\nonumber\\
&=&
\Omega
\int {\d^d \vec k \; \d \omega\over (2\pi)^{d+1}}
\ln X(\vec k, \omega).
\end{eqnarray}
Applying this to the one loop effective potential yields
\begin{eqnarray}
{\cal V}[\phi;\phi_0] &=&
\half F^2[\phi]  \; \tilde g_2^{-1}(\vec k=\vec 0,\omega=0)
- \half \A \int {\d^d \vec k \; \d \omega\over (2\pi)^{d+1}}
\ln \left[ D(\vec k,\omega) - {\delta F\over \delta\phi} \right]
- \half \A \int {\d^d \vec k \; \d \omega\over (2\pi)^{d+1}}
\ln \left[ D^\dagger(\vec k,\omega) - {\delta F\over \delta\phi}^\dagger
 \right]
\nonumber\\
&&+ \half \A \int {\d^d \vec k \; \d \omega\over (2\pi)^{d+1}}
\ln
\left[
\left( D^\dagger(\vec k,\omega)  - {\delta F\over  
\delta\phi}^\dagger \right)
\tilde g_2^{-1}(\vec k,\omega)
\left(  D(\vec k,\omega) - {\delta F\over \delta\phi} \right) +
F[\phi] {\delta^2 F\over\delta\phi\;\delta\phi}
\; \tilde g_2^{-1}(\vec k=\vec 0,\omega=0)
\right]
\nonumber\\
&&- \left( \phi \to \phi_0 \right)
+ O(\A^2).
\label{E:ep1}
\end{eqnarray}
(Note that $g_2$ plays two rather different roles above.) We now adopt
the simplifying {\em convention} that
\begin{equation}
\int \d^d \vec x\; \d t \;  g_2^{-1}(\vec x,t)
\; = \; 1
\; = \; \tilde g_2^{-1}(\vec k=0,\omega=0).
\end{equation}
This is only a {\em convention}, not an additional restriction on the
noise, since it only serves to give an absolute meaning to the
normalization of the amplitude $\A$.

With these conventions, the one-loop effective potential can be
written as
\begin{eqnarray}
{\cal V}[\phi;\phi_0] &=&
\half F^2[\phi]
- \half \A \int {\d^d \vec k \; \d \omega\over (2\pi)^{d+1}}
\ln \left[ D(\vec k,\omega) - {\delta F\over \delta\phi} \right]
- \half \A \int {\d^d \vec k \; \d \omega\over (2\pi)^{d+1}}
\ln \left[ D^\dagger(\vec k,\omega) - {\delta F\over \delta\phi}^\dagger 
\right]
\nonumber\\
&&+ \half \A \int {\d^d \vec k \; \d \omega\over (2\pi)^{d+1}}
\ln
\left[
\left( D^\dagger(\vec k,\omega)  - {\delta F\over  
\delta\phi}^\dagger \right)
\tilde g_2^{-1}(\vec k,\omega)
\left(  D(\vec k,\omega) - {\delta F\over \delta\phi} \right) +
F[\phi] {\delta^2 F\over\delta\phi\;\delta\phi}
\right]
\nonumber\\
&&- \left( \phi \to \phi_0 \right)
+ O(\A^2),
\label{E:ep2}
\end{eqnarray}
which can be recast into
\begin{eqnarray}
\label{E:general}
{\cal V}[\phi;\phi_0] &=&
\half F^2[\phi]
+ \half \A \int {\d^d \vec k \; \d \omega\over (2\pi)^{d+1}}
\ln
\left[ 1 + {\tilde g_2{}(\vec k,\omega)
F[\phi] {\delta^2 F\over\delta\phi\;\delta\phi}
\over
\left( D^\dagger(\vec k,\omega)  - {\delta F^\dagger \over \delta\phi}
\right)
\left( D(\vec k,\omega) - {\delta F\over \delta\phi} \right)}
\right]
\nonumber\\
&&- \left( \phi \to \phi_0 \right)
+ O(\A^2).
\end{eqnarray}
This formula is one of the central results of this paper. It shows
that noise induced fluctuations modify the zero-loop piece of the
potential in a way which is reminiscent of the situation in both
statistical and quantum field theory. For example, in QFT one
has~\cite{Weinberg-2,Rivers}:
\begin{eqnarray}
\label{E:general1}
{\cal V}_{QFT}[\phi;\phi_0] &=&
V(\phi)
+ \half \hbar \int {\d^d \vec k \; \d \omega\over (2\pi)^{d+1}}
\ln
\left[ 1 + { {\delta^2 V\over\delta\phi\;\delta\phi}
\over
\omega^2 + \vec k^2 + m^2
}
\right]
- \left( \phi \to \phi_0 \right)
+ O(\hbar^2).
\end{eqnarray}
We see in equation (\ref{E:general}) that the ground state structure
of the SPDE (which we will soon see is obtained by minimizing ${\cal
V}[\phi;\phi_0]$) depends on both the noise correlations and the
nonlinearities induced by the forcing term. We also see explicitly how
the noise amplitude is essential in the competition between
deterministic and stochastic effects.

The major difference between the effective potential for SPDEs and
QFT, lies in the fact that for SPDEs the scalar propagator of QFT is
replaced with a propagator which has a more complex structure for the
equivalent of the ``mass'' term. This difference owes to the causal
structure of SPDEs.

Notice also that for SPDEs one can naturally adapt the noise to be
both the source of fluctuations {\em and} the regulator to keep the
Feynman diagram expansion finite. This follows immediately by
inspection of (\ref{E:general}) which shows that the (momentum and
frequency dependent) noise shape function $\tilde g_2$ will affect the
momentum and frequency behavior of the one-loop integral. The
finiteness, divergence structure, and renormalizability of this
integral will very much depend on the functional form of $\tilde g_2$.
It is thus clear that we can use the noise shape function to regulate
the integral, if we wish.

\section{Interpretation}

The {\em physical interpretation} of the effective action and the
effective potential for SPDEs is considerably more subtle than that
for the more usual QFTs. The situation is complicated by the fact that
for a completely general SPDE it may not be meaningful to define a
physical energy. Even when the SPDE is sufficiently special so that
some physical notion of energy may be defined, the system may be
subject to dissipation: The {\em physical} energy need not be
conserved, even in the absence of noise. Thus the effective action and
effective potential for SPDEs are not related to the physical energy.
This means that {\em some} of the physical intuition built up from
QFTs may be misleading and it becomes important to reassess the notion
of effective action, and effective potential to see how much survives
in the SPDE context.

The great virtue of the effective action and effective potential in
QFT is that they contain all the information regarding the ground
state of the system and its fluctuations: From a knowledge of the
effective potential one can ascertain under what conditions the system
will display one degree of symmetry or another. It is essential that
most of these properties carry over to the case of SPDEs, otherwise
the effective action and effective potential would be mere
mathematical constructs without physical relevance.  Fortunately the
key features do in fact carry over: (1) the stationary points of the
effective action still correspond to stochastic expectation values of
the fields in the absence of an external current; (2) the effective
potential governs the probability that the {\em spacetime average} of
the field takes on specific values; (3) even when the notion of
physical energy is lacking we will see that there is a notion of
quasi-energy for SPDEs, with the quasi-energy being a measure of the
extent to which the system has been driven away from its
non-stochastic (zero-noise) configuration; and (4) the one-loop
effective action will be demonstrated to describe the (approximate)
probability for an initial field configuration to evolve into some
final (in the asymptotic sense) field configuration under the
influence of the stochastic noise.

\subsection{Equations of motion in the presence of fluctuations}

If one makes use of the definition of the effective action as a
  Legendre transform, it is easy to see that
\begin{equation}
{\delta \Gamma[\phi;\phi_0] \over \delta \phi} =  J[\phi],
\end{equation}
where $J[\phi]$ is that external current required in order that
\begin{equation}
\langle \phi[J] \rangle = \phi.
\end{equation}
In particular, by taking $J=0$
\begin{equation}
{\delta \Gamma[\phi;\phi_0] \over \delta \phi} = 0
\qquad \iff \qquad
\phi = \langle \phi[J=0] \rangle.
\end{equation}
Stationary points of the effective action occur at those (mean) field
configurations which are zero-external-current stochastic expectation
values of the fluctuating field. (Proof of this may be found for
instance on page 65 of Weinberg~\cite{Weinberg-2}). It is important to
realize that one never needs to invoke the notion of energy to obtain
this result. The QFT interpretation of this result, which we now see
extends to SPDEs, is that the effective action gives the equations of
motion once fluctuations (noise) are taken into account. (This is a
non-perturbative result, not limited to the one-loop approximation.)

\subsection{Probability distribution for the spacetime average field}

We have previously seen that the probability distribution for the
fluctuating field, considered as a function over spacetime, to take on
the value $\phi(\vec x,t)$ is given by the functional
\begin{equation}
\PPP[\phi] = \PP[D\phi-F[\phi]] \; \sqrt{\J \J^\dagger}.
\end{equation}
Now suppose we coarse-grain, by looking at the spacetime average of
the field $\phi$ as defined by
\begin{equation}
{\int_{\Omega_s\times T} \phi(\vec x,t) \; d^d\vec x \; dt
\over \Omega_s T},
\end{equation}
and ask: what is the probability that this spacetime average take on a
specific numerical value $\bar\phi$?  (For definiteness we impose
periodic boundary conditions in space $\Omega_s$ and time $T$ and
interpret $\Omega_s T$ as the volume of the spacetime box. This has
the technical advantage that the partition function $Z[J]$ is then
needed only for sources $J$ that are strictly independent of space and
time.)

The probability we are interested in is easily calculated to be
\begin{eqnarray}
\hbox{Prob}\left( \int_{\Omega_s\times T} \d^d \vec x \; \d t \;
\phi(\vec x,t) =
\bar\phi \; \Omega_s T\right)
&=&
\int ({\cal D} \phi) \; \PPP[\phi] \;
\delta\left(
\int_{\Omega_s \times T} \d^d \vec x \; \d t \;
 \phi(\vec x,t) - \bar\phi \; \Omega_s T
\right)
\\
&=&
\int ({\cal D} \phi) \int d\lambda \; \PPP[\phi] \;
\exp\left(i\lambda\left[
\int_{\Omega_s\times T} \d^d \vec x \; \d t \;
\phi(\vec x,t) - \bar\phi \; \Omega_s T
\right]\right)
\\
&=&
\int d\lambda \; Z[J(x)=i\lambda] \; \exp(-i\lambda\Omega_s T\;\bar\phi).
\end{eqnarray}
We now take the limit as $\Omega_s T$ becomes very large, and apply
the method of stationary phase. By definition we have
\begin{eqnarray}
Z[J(x)=i\lambda] =
\exp\left[
\Omega_s T
\{ i\lambda \phi(\lambda) - {\cal V}[\phi(\lambda);\phi_0]
/{\cal A} \}
\right]
\end{eqnarray}
with the  subsidiary condition
\begin{equation}
\left.
\frac{\delta {\cal V}[\phi;\phi_0]}{\delta \phi}
\right|_{\phi(\lambda)} = i \lambda {\cal A}.
\end{equation}
It is easy to demonstrate that
\begin{eqnarray}
\hbox{Prob}\left(
\int_{\Omega_s\times T} \d^d \vec x \; \d t \;
\phi(\vec x,t) = \bar\phi \; \Omega_s T
\right)
&\propto&
\exp\left(
-\Omega_s T\left[
{{\cal V}[\bar\phi;\phi_0]\over{\cal A}} +
O\left({1\over\Omega_s T}\right)
\right]
\right).
\end{eqnarray}
Thus the effective potential governs the probability distribution of
the spacetime average of the fluctuating field. Minima of the
effective potential correspond to maxima of the probability density of
the spacetime average field. The way we have set up the argument
applies equally well to QFTs and SPDEs and makes no reference to the
notion of physical energy. (This result is non-perturbative but
approximate---it is not limited to one loop. If we take either the
infinite volume or infinite time limits, then with probability one,
the spacetime average field must equal one of the minima of the
effective potential.)

\subsection{Action and quasi-energy for SPDEs}

Even though the physical energy may not be defined for arbitrary
SPDEs, we nevertheless can demonstrate that there always exists a
positive-semi-definite functional of field configurations, the
tree-level action, and a related ``quasi-energy'', whose minima
correspond to maxima of the probability distribution of field
configurations.

{From} the way the functional formalism has been set up, we can always
define and calculate an effective action and an effective potential
even if the underlying non-stochastic version of the partial
differential equation does not arise from a Lagrangian formulation. We
have already seen that the effective action has a natural
interpretation in terms of the equations of motion once fluctuations
are taken into account, and that the effective potential governs
fluctuations in the spacetime average of the field. We now go one step
further: We distinguish two concepts of ``energy'', the ``true
physical energy'' and the ``quasi-energy'', and show that even if the
physical energy is undefined (or possibly not useful due to
dissipative effects), the quasi-energy is still a useful measure of
the extent to which fluctuations modify the non-stochastic equations
of motion.  We start from our general SPDE (\ref{E:eom0})
\begin{equation}
\label{E:noisy}
D \phi = F[\phi] + \eta,
\end{equation}
and its non-stochastic version
\begin{equation}
\label{E:zero-noise}
D\phi = F[\phi].
\end{equation}
Sometimes this non-stochastic partial differential equation will arise
from some Lagrangian, often it will not.

Even if the PDE $(D\phi = F[\phi])$ does not arise from a Lagrangian,
the results of this paper demonstrate that it is always possible to
assign a tree-level action to the stochastic system:
\begin{equation}
{\cal S}_{\mathrm classical}  = \half \int \int dx\,dy\,
(D \phi - F[\phi]) g_2^{-1} (D \phi - F[\phi]) \geq 0.
\end{equation}
This classical action is positive semidefinite, and has minima (which
are equal to zero) at field configurations that satisfy the zero-noise
equations of motion.  This is most obvious for white noise, when the
action is a perfect square, but the result is general.  The noise
two-point correlation function, [being an (infinite dimensional)
covariance matrix], is by definition positive definite. Therefore its
inverse is also positive definite and similarly the (infinite
dimensional) matrix $g_2^{-1}$ is a positive definite operator.  Thus
this classical action (the generalized Onsager--Machlup
action~\cite{Onsager-Machlup}) is always greater than or equal to
zero.

The classical action thus measures the extent to which a given field
configuration fails to satisfy the zero-noise equations of motion; the
measure of the deviation being weighted by the {\em shape} of the
noise correlations. (In fact, if the amplitude ${\cal A}$ of the noise
is set to zero, the action is identically equal to zero.)

We now define the quasi-energy by
\begin{equation}
\label{quasiHam}
{\cal S}[\phi] = \int E_{\mathrm quasi}[\phi] \; dt.
\end{equation}
We justify calling this object the quasi-energy by the fact that if we
treat it as a Hamiltonian functional, and put the resulting object
into the partition function of an equilibrium statistical field
theory, we get the generating functional for all the correlation
functions (ignoring ghost Jacobians for the moment). Explicitly, we
can write
\begin{equation}
E_{\mathrm quasi}[\phi]  =
\half \int \int d^d \vec x \; d^d \vec y \; dt' \;
(D \phi - F[\phi]) g_2^{-1} (D \phi - F[\phi]).
\end{equation}
Note that the quasi-energy depends both on the PDE and on the shape of
the noise correlation function. If the amplitude ${\cal A}$ of the
noise is set to zero, this quasi-energy is conserved and is exactly
equal to zero. This quasi-energy can be thought of as a nonlinear
generalization of the Onsager--Machlup ``energy'' to arbitrary
Gaussian noise\footnote{The particular label one chooses to apply to
this quantity is not important as long as one bears carefully in mind
that this ``energy'' need not be the physical energy.}.

If we now restrict ourselves to homogeneous and static fields, and
consider the effective potential as defined above, then by the
procedures used in quantum and stochastic field theories, the effective
potential (multiplied by the volume of space) is the stochastic
expectation value of the quasi-energy $\langle E_{\mathrm
quasi}[\phi]\rangle$ in the {\em presence} of the noise induced
fluctuations, and subject to the constraint $\langle\phi\rangle =
\phi$. A proof of this result is provided on pages 72--73 of
Weinberg~\cite{Weinberg-2}. Though that proof is phrased in a
Lorentzian-signature QFT language, it readily carries over to
Euclidean-signature equilibrium statistical field theory.  Once the
physical energy is replaced by the quasi-energy, the proof can be
extended to SPDEs as well.

We have been able to show that minima of the effective potential also
minimize the quasi-energy, and therefore the noise-induced deviations
from the zero-noise equations of motion.

\subsection{Transition probabilities}

What is the probability that a certain initial field configuration
$\phi_i(\vec x)$ at time $t_i$ evolves into a final field
configuration $\phi_f(\vec x)$ at time $t_f$? We have already
developed the appropriate machinery to address this question. Indeed
\begin{eqnarray}
\hbox{Prob}(\phi_f(\vec x),t_f; \phi_i(\vec x), t_i)
&\propto&
\int ({\cal D}\eta) \; \PP[\eta] \;
\delta[\phi_{\mathrm soln}(\vec x, t_i;\eta) - \phi_i(\vec x)]\;
\delta[\phi_{\mathrm soln}(\vec x, t_f;\eta) - \phi_f(\vec x)]
\nonumber\\
&\propto&
\int ({\cal D}\eta) \; ({\cal D} \phi) \; \PP[\eta] \;
\delta[\phi_{\mathrm soln}(\vec x, t;\eta) - \phi(\vec x,t)]\;
\delta[\phi(\vec x, t_i) - \phi_i(\vec x)]\;
\delta[\phi(\vec x, t_f) - \phi_f(\vec x)]
\nonumber\\
&\propto&
\int ({\cal D}\eta) \; ({\cal D} \phi) \; \PP[\eta] \;
\delta[ D\phi - F[\phi] - \eta] \; \sqrt{ {\cal J J}^\dagger } \;
\delta[\phi(\vec x, t_i) - \phi_i(\vec x)]\;
\delta[\phi(\vec x, t_f) - \phi_f(\vec x)]
\nonumber\\
&\propto&
\int ({\cal D} \phi) \;
\PP[D\phi - F[\phi]] \; \sqrt{ {\cal J J}^\dagger } \;
\delta[\phi(\vec x, t_i) - \phi_i(\vec x)]\;
\delta[\phi(\vec x, t_f) - \phi_f(\vec x)]
\nonumber\\
&\propto&
\int ({\cal D} \phi) \;
\PPP[\phi] \;
\delta[\phi(\vec x, t_i) - \phi_i(\vec x)]\;
\delta[\phi(\vec x, t_f) - \phi_f(\vec x)]
\nonumber\\
&\propto&
\int ({\cal D} \phi) \;
\exp\left(-{\cal S}[\phi]/{\cal A}\right) \;
\sqrt{ {\cal J J}^\dagger } \;
\delta[\phi(\vec x, t_i) - \phi_i(\vec x)]\;
\delta[\phi(\vec x, t_f) - \phi_f(\vec x)]
\nonumber\\
&\propto&
\int_{\phi(\vec x, t_i)=\phi_i(\vec x)}^{\phi(\vec x,  
t_f)=\phi_f(\vec x)}
\; ({\cal D} \phi) \;
\exp\left(-{\cal S}[\phi]/{\cal A}\right) \;
\sqrt{ {\cal J J}^\dagger }.
\end{eqnarray}
This is formally identical to the formula usually encountered in
equilibrium statistical field theory, and everything so far is
non-perturbatively correct.

Now take a saddle point approximation: Find an interpolating field
$\phi_{\mathrm int}(\vec x,t)$ that minimizes ${\cal S}[\phi]$ and
interpolates from $\phi_i(\vec x)$ to $\phi_f(\vec x)$. Perform the
Gaussian integral about the saddle point. Then by definition of the
one-loop effective action
\begin{equation}
\hbox{Prob}(\phi_f(\vec x),t_f; \phi_i(\vec x), t_i)
\approx
\exp\left[-{\Gamma[\phi_{\mathrm int}]\over{\cal A}}\right].
\end{equation}
This is only a one-loop result, but it demonstrates that
the effective action for SPDEs inherits many of the important features
of the effective action for QFTs.

\subsection{Summary}

{From} the above, we see that the effective action and effective
potential for SPDEs exhibit many of the key features of the effective
action and effective potential of QFTs. This is important because it
guarantees that not only is it relatively easy to calculate the
one-loop effective potential, but also it is useful to do so: As is
the case for QFTs, minima of the effective potential for SPDEs provide
information about expectation values of the fields. The effective
action also provides information about fluctuations in spacetime
averaged fields, it gives information about the noise-induced
deviations from the non-stochastic equations of motion, and it governs
the transition probabilities whereby initial field configurations
evolve to final field configurations. Thus, both the effective
potential and the effective action are as useful for SPDEs as it is
for QFTs.

Furthermore, as demonstrated in recent work by Alexander and
Eyink~\cite{Eyink1,Eyink2,Eyink3}, the effective potential is also a
useful tool in a strong noise regime far from equilibrium.  The major
difference between those papers and our own formalism is that they
work within the MSR approach. They also focus on strong noise regimes,
while we emphasize that for many purposes a one-loop calculation is
both computationally efficient and quite sufficient to extract many
key features of the physics of the system. The two approaches are
complementary, and where they overlap, they are in complete agreement.

\section{Discussion}

In this paper we have developed a general and powerful formalism
applicable to arbitrary SPDEs. We have shown how to convert {\em
arbitrary} correlation functions associated with {\em arbitrary} SPDEs
into functional integrals. (And for this first step the noise does not
have to be Gaussian.) For Gaussian noise (not necessarily
translation invariant) we have carried the formalism further, setting
up the basic ingredients needed for Feynman diagram expansions with
the noise amplitude serving as the loop-counting parameter, and
defining a non-perturbative effective action in analogy with QFT.

We hope to have convinced the reader that the ``direct approach''
developed in this paper is both useful and complementary to the more
traditional MSR formalism \cite{MSR,De-Dominicis-Peliti,Zinn-Justin}.
Some questions can more profitably be asked and answered in this
``direct'' formalism. For instance, the fact that the noise amplitude
is the loop-counting parameter is easy to establish in this ``direct''
formalism, but appears to have no analog result in the MSR formalism.
The effective action gives rise naturally to the concept of an
effective potential, a powerful construct well known and studied
within the QFT context, where it serves to classify and compute ground
states and allows one to investigate symmetry properties and patterns
of symmetry breaking (both spontaneous and dynamic). An analogous
construct can also be defined and calculated for stochastic field
theories based on SPDEs, and we have done so in this work. However,
for arbitrary SPDEs, such as those contemplated here, the notion of
ground state and effective potential must be approached with extra
care and their physical interpretation clarified.  We have taken pains
to do so, establishing that the minimum of the construct we call the
effective potential corresponds to solving the full equations of
motion (for homogeneous and static field configurations) in the
presence of noise.

We feel that the most interesting result of this analysis is a general
formula for the one-loop effective potential for {\em any} SPDEs
subject to translation-invariant Gaussian noise.  This is still an
extremely broad class of problems, and in a pair of companion papers
we will specialize this analysis to two particular cases. First, we
discuss the noisy Burgers equation (KPZ equation), where the effective
potential approach immediately leads us to such interesting
observations as the existence of dynamical symmetry breaking (DSB) and
the Coleman--Weinberg mechanism~\cite{HMPV-kpz}. Second, we discuss
the reaction--diffusion-decay system, and explicitly calculate the
renormalized effective potential for 1, 2, and 3 spatial
dimensions~\cite{HMPV-rdd}. These are issues that are extremely
difficult to address using the MSR approach.

\section*{Acknowledgments}

In Spain, this work was supported by the Spanish Ministry of Education
and Culture and the Spanish Ministry of Defense (DH and JPM). In the
USA, support was provided by the US Department of Energy (CMP and MV).
This research is supported in part by the Department of Energy under
contract W-7405-ENG-36 (CMP).  Additionally, MV wishes to acknowledge
support from the Spanish Ministry of Education and Culture through the
Sabbatical Programme, to recognize the kind hospitality provided by
LAEFF (Laboratorio de \Astrofisica\ Espacial y \Fisica\ Fundamental;
Madrid, Spain), and to thank Victoria University (Te Whare Wananga o
te Upoko o te Ika a Maui; Wellington, New Zealand) for hospitality
during final stages of this work.

\appendix
\section{Jacobian functional determinant}

The Jacobian functional determinant is often (but not always) field
independent, and can often (but not always) be discarded. In this
Appendix we explore this issue in more detail.

\subsection{Causality: Retarded Green functions}

This discussion is a generalization of Rivers~\cite{Rivers},
pp. 155--156. There are also relevant comments in De Dominicis and
Peliti~\cite{De-Dominicis-Peliti}, Appendix B part C (pp.
370--371). See also the footnote on page 214 of Frisch~\cite{Frisch},
and the discussion in Zinn--Justin (pp. 372--373~\cite{Zinn-Justin}).
We are interested in evaluating
\begin{equation}
{\cal J} \equiv \det\left( D - {\delta F\over\delta\phi} \right).
\end{equation}
To proceed we make some specific assumptions about the form of
$D$. Let us confine attentions to the class of differential operators
\begin{equation}
D_n \equiv {\partial^n\over\partial t^n} - D_0(\vec \nabla).
\end{equation}
(If we take $D_0 = \vec \nabla^2$ then $D_1$ is the diffusion operator
while $D_2$ is the wave operator, so this class of differential
operators is still broad enough to cover almost everything of
physical interest.) Now write
\begin{eqnarray}
{\cal J}_n
&\equiv& \det\left( \partial_t^n - D_0 - {\delta F\over\delta\phi}  
\right)
\\
&=&\det\left( \partial_t^n \right) \;
\det\left[ I - G_n \left(D_0 + {\delta F\over\delta\phi} \right) \right]
\\
&=&\det\left( \partial_t^n \right) \:
\exp\Tr\ln\left[ I - G_n \left(D_0 + {\delta F\over\delta\phi}  
\right) \right]
\\
&=&\det\left( \partial_t^n \right) \;
\exp\left\{-\sum_{m=1}^\infty \Tr\left( {1\over m}
\left[G_n \left(D_0 + {\delta F\over\delta\phi} \right) \right]^m \right)
\right\}.
\end{eqnarray}
Here $G_n$ is the retarded Green function corresponding to
$\partial_t^n$. Explicitly
\begin{equation}
G_n(t,t') = {(t-t')^{n-1}\over(n-1)!} \; \Theta(t-t').
\end{equation}
One can easily check that this is a Green function by computing,
for $n>1$,
\begin{equation}
\partial_t G_n(t,t') =
{(t-t')^{n-2}\over(n-2)!} \; \Theta(t-t') =
G_{n-1}(t,t'),
\end{equation}
and noting that
\begin{equation}
\partial_t G_1(t,t') =  \delta(t-t').
\end{equation}
Finally, the retarded nature of the Green function is due to the
 presence of the Heaviside step function.

The traces $\Tr$ in the formula for the Jacobian ${\cal J}_n$ are
 spacetime traces.  We may write this as $\Tr = \tr_{\mathrm time}\;
 \tr_{\mathrm space}$, and concentrate (for now) only on the trace
 over time $\tr_{\mathrm time}$.  For $n>1$ it is easy to see that
 $\tr_{\mathrm time}(G_n)=0$, and in fact that for all $m>0$,
 $\tr_{\mathrm time}([G_n]^m)=0$.  To generalize this argument to the
 spacetime trace we need to make the assumption that $F[\phi(\vec
 x,t)]$ does not explicitly contain any time derivatives.  If this is
 the case, we can write
\begin{equation}
{\delta F[\phi(\vec x,t)]\over\delta\phi(\vec y,t')} =
\delta(t-t') \; {\delta F[\phi(\vec x,t)]\over\delta\phi(\vec y,t)}.
\end{equation}
This now implies that for the spacetime trace ($n>1$; $m>0$)
\begin{equation}
\Tr\left(
\left[G_n \left\{D_0+{\delta F[\phi]\over\delta\phi}\right\}\right]^m 
\right) = 0.
\end{equation}
Thus the retarded nature of the Green function, causes all the trace
terms to vanish and we have the exact result that for $n>1$ and
$F[\phi]$ not containing time derivatives
\begin{eqnarray}
{\cal J}_n
&\equiv& \det\left( \partial_t^n - D_0 - {\delta F\over\delta\phi}  
\right)
\\
&=&\det\left( \partial_t^n \right).
\end{eqnarray}
This means that the functional determinant is simply a field
independent constant. It is therefore irrelevant and may be
discarded. (In particular, for $n=2$, the stochastic wave equation,
one never has to evaluate the functional determinant.)  ({\em Aside:}
This argument also works provided $n > 1 +$ [the highest order of time
derivatives occurring in $F[\phi] $].  Proving this is an easy
exercise.) The partition function (characteristic functional) is now,
for $n>1$:
\begin{eqnarray}
Z[J] &\propto& \int ({\cal D} \phi)\;
         \exp\left[ -\half
              \int \int (D_n\phi-F[\phi])  G_\eta^{-1} (D_n\phi-F[\phi])
             \right] \;
        \exp\left( \int J \phi \right).
\end{eqnarray}
For $n=1$ the situation is almost as good. First note that
\begin{eqnarray}
(G_1)^2(t,t')
&=& \int \d \bar t \; G_1(t,\bar t) \; G_1(\bar t, t')
\\
&=& \int \ d \bar t \; \Theta(t-\bar t) \; \Theta(\bar t-t')
\\
&=& (t-t') \;  \Theta(t-t').
\end{eqnarray}
Thus $\tr([G_1]^2)=0$, and it is easy to show that for $m>1$, $(G_1)^m
= G_m$ so that $\tr([G_1]^m)=0$. The only term that survives is
$\tr(G_1) = \Theta(0)$. But $\Theta(0)$ is ill-defined and must be
specified by some particular prescription.  The prescription which is
most useful in this context is the symmetric one wherein $\Theta(0)$
is non zero and equals ${1\over2}$. This may be justified by a
limiting procedure as described, for example, in the text by
Zinn--Justin~\cite{Zinn-Justin} (Chapter 4, pp. 69--70.)  This
symmetric prescription is equivalent to adopting the {\em Stratonovich
calculus} for stochastic equations. Choosing $\Theta(0)=0$ is
equivalent to the {\em Ito calculus}. The Ito calculus simplifies the
Jacobian determinant (to unity) at the cost of destroying equivariance
under field redefinitions (the Ito calculus explicitly breaks
coordinate invariance in field space). See, for instance,
Eyink~\cite{Eyink1}, or Zinn--Justin~\cite{Zinn-Justin}.  We will
stick with the symmetric prescription (Stratonovich calculus) for this
paper, though suitable modifications for the Ito calculus are
straightforward if at times tricky (the loss of reparameterization
invariance under field redefinitions implies that all arguments
involving a change of variables must be carefully re-assessed).

Now for $n=1$ only one of the trace terms in the functional
determinant survives and we have, (with the assumption that $F[\phi]$
contains no time derivatives)
\begin{eqnarray}
\Tr\left[ G_1 {\delta F[\phi(x)]\over\delta\phi(y)} \right]
&=&
\Tr\left[ G_1 {\delta F[\phi(\vec x,t)]\over\delta\phi(\vec y,t)} \right]
\nonumber\\
&=&
\Theta(0) \int \d t \;
\tr_{\mathrm space}
\left[ {\delta F[\phi(\vec x,t)]\over\delta\phi(\vec y,t)} \right]
\nonumber\\
&=&
\Theta(0) \;
\Tr\left[ {\delta F[\phi(\vec x,t)]\over\delta\phi(\vec y,t)} \right].
\end{eqnarray}
This implies
\begin{eqnarray}
{\cal J}_1
&\equiv&
\det\left( \partial_t - D_0 - {\delta F\over\delta\phi} \right)
\\
&=&\det\left( \partial_t \right) \;
\exp\left\{-\Theta(0)\; \Tr
\left[D_0 + {\delta F[\phi(\vec x)]\over\delta\phi(\vec y)} \right]
\right\}
\\
&=&\det\left( \partial_t \right) \;
\exp\left\{-\half\Tr[D_0]\right\} \;
\exp\left\{
-\half
\Tr\left[{\delta F[\phi(\vec x)]\over\delta\phi(\vec y)}\right]
\right\}.
\end{eqnarray}
The first two factors in the last line are field independent and so
 may be discarded with the result that
\begin{equation}
{\cal J}_1 \propto
\exp \left\{-\half
\Tr\left[{\delta F[\phi(\vec x)]\over\delta\phi(\vec y)}\right]
\right\}.
\end{equation}
The partition function (characteristic functional) is now
\begin{eqnarray}
Z[J] &\propto& \int ({\cal D} \phi)\;
         \exp\left[ -\half
                   \int \int (D_1\phi-F[\phi])  G_\eta^{-1}  
(D_1\phi-F[\phi])
             \right] \;
\\
&&\qquad
\exp\left(-\half
\Tr\left[{\delta F[\phi(\vec x)]\over\delta\phi(\vec y)} \right]
\right)
\; \exp\left( \int J \phi \right).
\end{eqnarray}
This means that for stochastic differential equations that are first
order in time, the functional determinant must be kept. There are
specific choices of the nonlinear driving term $F[\phi]$ that lead to
even further simplifications.

\subsection{Jacobian functional determinant for local driving  
forces}

Suppose that $F[\phi(x)]$ is a {\em local}\, functional of the field
$\phi$. This implies that there exists a local function ${\cal
F}(\phi,\vec \nabla)$ such that
\begin{equation}
{\delta F[\phi(\vec x,t)] \over \delta\phi(\vec y, t')} =
{\cal F}(\phi(\vec x,t),\vec \nabla) \;
\delta(t-t') \; \delta(\vec x - \vec y).
\end{equation}
Evaluating the functional determinant now gives
\begin{equation}
{\cal J} = \exp\left( -\half \delta^d(\vec 0 )
\int \d t \; \d^d \vec x \; {\cal F} \right).
\end{equation}
Insofar as we trust the formal result $\delta^d(\vec 0)=0$ (see, for
example, \cite{Zinn-Justin}) we can discard the functional determinant
as an irrelevant constant\footnote{
This formal result is somewhat of a contentious issue, and we have
found that it is often more useful and safer to either prove that the
Jacobian is a field-independent constant~\cite{HMPV-kpz}, or to carry
the Jacobian along for the whole calculation~\cite{HMPV-rdd}.
}.

For more general driving terms $F[\phi]$ one must keep the functional
determinant. Nevertheless it is clear that for large classes of
stochastic partial differential equations, including many of the most
important and interesting cases, the Jacobian can be safely ignored.

For differential operators $D$ that are not of the form $D_n$
discussed above, or driving forces $F$ more complicated than those
discussed above, one has to use other means of evaluating the
functional determinant.

For the noisy Burgers equation (KPZ system) the functional determinant
can be shown to be a field independent constant that can be discarded.
A proof of this will be provided in~\cite{HMPV-kpz}. For the
reaction-diffusion-decay system on the other hand, we find it more
convenient to explicitly keep the Jacobian
determinant~\cite{HMPV-rdd}.

\subsection{Jacobian functional determinant via Feynman diagrams}

Let us now suppose that $D=D_n$, as discussed in the first section of
this Appendix. Then $D(\vec k,\omega) = (-i\omega)^n - D_0(\vec k)$
and\ the ghost propagator is
\begin{eqnarray}
G_{\mathrm ghost}(\vec k,\omega) &=& {1\over (-i\omega)^n - D_0(\vec k)},
\end{eqnarray}
where the retarded nature of the Green function now implies that all
poles in the $\omega$ plane occur in the lower half of this plane.
For each ghost loop (assuming $F[\phi]$ contains no time derivatives)
we must perform an integral over both frequency and momenta of the
type
\begin{equation}
{\cal I}_{n}{}^{m} \equiv \int
{ \d\omega \; \d^d \vec k \; P(\vec k, \vec k_i)\over (2\pi)^{d+1} \;
\Pi_{i=1}^m
\{[-i(\omega-\omega_i)]^n - D_0(\vec k -\vec k_i) \}},
\end{equation}
where all the poles (in $\omega$) lie in the lower half-plane and the
$\omega_i$ and $\vec k_i$ are linear combinations of the momenta
flowing into the ghost loop. (The function $P(\vec k, \vec k_i)$ is
some possibly complicated function of the momenta, typically a
polynomial, derived from $\delta F/\delta\phi$. There is also a set of
external legs (derived from $\delta F/\delta\phi$) attached to each
vertex of the ghost loop, but we do not need to know the detailed
structure of these vertices to derive the expression above.)

Since all the poles are known to lie in the lower half plane the
contour of integration can be pushed out to infinity in the upper
half-plane via the replacement $\omega \to \omega +i\Lambda$ ($\Lambda
> 0$), without changing the value of the integral. Thus we can write
\begin{equation}
\forall \Lambda > 0: \qquad {\cal I}_{n}{}^{m} =
\int { \d \omega \; \d^d \vec k \; P(\vec k, \vec k_i) \over  
(2\pi)^{d+1} \;
\Pi_{i=1}^m
\{[-i(\omega-\omega_i)-\Lambda]^n - D_0(\vec k -\vec k_i) \}}.
\end{equation}
Now take the limit $\Lambda\to+\infty$ to deduce ${\cal I}_{n}{}^{m} =
0$.

The only place that this argument fails is when the $\omega$ integral
does not converge. This happens only for $n=1$ (first-order in time)
and $m=1$ (tadpole diagram) in which case we need to consider
\begin{equation}
{\cal I}_{1}{}^{1} \equiv
\int { \d \omega \; \d^d \vec k \; P(\vec k) \over
(2\pi)^{d+1} \{-i\omega - D_0(\vec k) \}}.
\end{equation}
This already reproduces the key results of the previous section: The
functional determinant can be ignored for $n>1$ and for $n=1$ it
collapses to a single term. Performing the $\omega$ integral for this
remaining term we see
\begin{eqnarray}
{\cal I}_{1}{}^{1}
&=& \int {\d^d \vec k\over (2\pi)^{d+1}} \;
i \; \ln[-i\omega - D_0(\vec k)]|_{\omega=-\infty}^{\omega=+\infty}
\\
&=& \int {\d^d \vec k\over (2\pi)^{d+1} }\; i \; (i\pi)
\\
&=& -\half {\int \d^d \vec k \over  (2\pi)^d }
\\
&=& -\half \; \delta^d(\vec 0).
\end{eqnarray}
We conclude then, that the tadpole ghost diagram exactly reproduces
the $\exp( -{1\over2} \Tr(\delta F/\delta \phi))$ obtained by other
means in the first section of this Appendix. In fact, Faddeev--Popov
ghost techniques are in complete agreement with direct calculations of
the functional determinant. This ghost-based analysis also makes clear
why things are different in QFT. If one uses the Feynman propagator
instead of the retarded propagator, there are poles on both sides of
the real line and one cannot push the path of integration out to
$+i\infty$.

\section{Equations of motion in the presence of noise}
\subsection{Effect of adding a small decay term}

We start with the zero-loop equations of motion for the SPDE
\begin{equation}
\left( D^\dagger - {\delta F\over \delta\phi}^\dagger \right)
\int g_2^{-1} \left(D\phi-F[\phi] \right) =J.
\end{equation}
In particular, for zero external source ($J=0$) any solution of the
non-stochastic bare equations of motion, $D\phi-F[\phi] =0$, is also a
solution of the zero loop equations of motion. (Zero loops {\em
almost} corresponds to setting the noise amplitude to zero and
reducing the SPDE to its non-stochastic analog.) But there is a risk
that the zero-loop equations may have {\em more} solutions than the
non-stochastic bare equations. This potential problem arises if the
operator $D^\dagger - ({\delta F^\dagger/ \delta\phi})$ is singular
(so that it has a non-trivial null space [kernel]). If this operator
is singular then there will be many different fields $\phi(\vec x,t)$
that correspond to a given $J$, making the whole Legendre transform
procedure invalid.

The best way to fix this is to add a small decay term in the system
and then take the limit as the decay term vanishes.  Specifically,
take
\begin{equation}
F[\phi] \to F[\phi] - \epsilon \phi.
\end{equation}
This perturbed system has zero-loop equations given by
\begin{equation}
\left( D^\dagger - {\delta F\over \delta\phi}^\dagger + \epsilon I  
\right)
\int g_2^{-1} \left(D\phi-F[\phi]+\epsilon \phi \right) =J.
\end{equation}
Even if $D^\dagger - ({\delta F^\dagger/\delta\phi})$ is singular, the
perturbed operator will not be, and so the perturbed equations of
motion will have a unique solution $\phi_{\mathrm soln}(J,\epsilon)$.
It is appropriate to take the Legendre transform using this unique
solution and consider the limit $\epsilon\to0$ at the end of the
calculation. If $D^\dagger - {\delta F^\dagger/\delta\phi}$ is
non-singular this does not change anything. If $D^\dagger - {\delta
F/\delta\phi}^\dagger$ is singular this procedure provides a
prescription for defining a unique solution to the zero-loop
equations.

The above complication is not peculiar to SPDEs and their associated
non-quantum field theories; the same sort of behaviour also occurs in
ordinary QFTs. For example in QED, $J=0$ corresponds to arbitrary
constant electromagnetic field. In order to assure that $J=0$ has the
unique solution $F=0$, the easiest thing to do is to add a small
photon mass.

\subsection{A vanishing theorem}

Suppose $D^\dagger - {\delta F/\delta\phi}^\dagger$ is non-singular,
then the zero-loop equations of motion for $J=0$ are equivalent to the
classical equations $D \phi- F[\phi]=0$.  The effective action at
zero-loop order, evaluated on solutions of the zero-loop equations of
motion is  exactly zero.

Not only that, in fact, the one-loop effective action evaluated on
solutions of the zero-loop equations of motion is also exactly zero.
This happens due to the explicit occurrence of $D\phi-F[\phi]$ in the
one-loop contribution to the effective action, [see (\ref{E:Gamma-1})
or (\ref{E:Gamma-2})], so that for solutions of the zero-loop
equations of motion there is an exact cancellation between the
Jacobian and the fluctuation operator $S_2$.

On the other hand, if $D^\dagger - {\delta F/\delta\phi}^\dagger$ is
singular, just perturb the system with a small amount of $\epsilon$
decay. The previous argument goes through for $\epsilon\neq0$.
(Technically as long as $\epsilon$ is not an eigenvalue of $D^\dagger
- {\delta F/\delta\phi}^\dagger$.) Taking the limit $\epsilon\to 0$
justifies the extension of the vanishing result to the singular case.

Now consider solutions of the
one-loop equations of motion. These one-loop equations of motion are
of the form
\begin{equation}
\left( D^\dagger - {\delta F\over \delta\phi}^\dagger + \epsilon I\right)
g_2^{-1} \left(D\phi-F[\phi] +\epsilon\phi\right) =O(\A),
\end{equation}
with the RHS being a complicated expression. Nevertheless, we do not
need to know exactly what this term is to deduce that evaluated
at solutions of these equations of motion $\Gamma[\phi_{\mathrm soln}]
= 0 + O(\A^2)$.

This vanishing of the effective action at solutions of the one-loop
equations of motion provides a useful consistency check on specific
calculations. The underlying reason for this vanishing theorem is
most easily addressed in the MSR formalism. In fact, it can be shown that
\begin{equation}
\Gamma[\phi;\phi_0] =
\half \int \int
\left\{ (D_{\mathrm eff}\phi-F_{\mathrm eff} [\phi])
\; g_2^{-1} \;
(D_{\mathrm eff} \phi-F_{\mathrm eff}[\phi]) \right\}
\d^d \vec x \; \d t \; \d^d \vec y \; \d t',
\end{equation}
with $D_{\mathrm eff}$ and $F_{\mathrm eff}$ are some effective
differential operator and effective driving force appropriate to
the fully interacting theory.



\begin{references}
\bibitem{Textbooks} The number of textbooks on quantum field theory
has grown explosively over the last few years. Apart
from~\cite{Zinn-Justin,Weinberg-1,Weinberg-2,Rivers} there are now
textbooks available at almost every level of detail and clarity, from
from~\cite{Bjorken-Drell-1} to~\cite{Le-Bellac}.
\end{references}
\end{document}